\pdfoutput=1
\documentclass[11pt,a4paper]{article}
\usepackage{jheppub}
\usepackage[caption=false]{subfig}
\usepackage{graphicx}
\usepackage{hyperref}
\usepackage{multirow}
\usepackage{color}

\newcommand{\beq}{\begin{eqnarray}}
\newcommand{\eeq}{\end{eqnarray}}
\newcommand{\non}{\nonumber\\}

\newcommand{\p}{\partial}
\newcommand{\Tr}{\mathop{\rm Tr}}

\begin{document}

\title{Topological solitons in the supersymmetric Skyrme model}

\author{Sven Bjarke Gudnason,${}^{1}$}
\author{Muneto Nitta${}^2$ and}
\author{Shin Sasaki${}^3$}
\affiliation{${}^1$Institute of Modern Physics, Chinese Academy of Sciences,
  Lanzhou 730000, China}
\affiliation{${}^2$Department of Physics, and Research and
    Education Center for Natural Sciences, Keio University, Hiyoshi
    4-1-1, Yokohama, Kanagawa 223-8521, Japan}
\affiliation{${}^3$Department of Physics, Kitasato University
Sagamihara 252-0373, Japan}
\emailAdd{bjarke(at)impcas.ac.cn}
\emailAdd{nitta(at)phys-h.keio.ac.jp}
\emailAdd{shin-s(at)kitasato-u.ac.jp}

\abstract{
A supersymmetric extension of the Skyrme model was obtained recently,
which consists of only the Skyrme term in the Nambu-Goldstone (pion) 
sector complemented by the same number of 
quasi-Nambu-Goldstone bosons.
Scherk-Schwarz dimensional reduction yields 
a kinetic term in three or lower dimensions 
and a potential term in two dimensions, preserving supersymmetry. 
Euclidean solitons (instantons) are constructed in the supersymmetric Skyrme
model. In four dimensions, the soliton is an instanton first found by
Speight. 
Scherk-Schwarz dimensional reduction is then performed once
to get a 3-dimensional theory in which a 3d Skyrmion-instanton is found and
then once more to get a 2d theory in which a 2d vortex-instanton 
is obtained.  
Although the last one is a global vortex 
it has finite action in contrast to conventional theory. 
All of them are non-BPS states breaking all supersymmetries.
}

\keywords{Supersymmetry, Skyrmions, solitons}

\maketitle

%%%%%%%%%%%%%%%%%%%%%%%%%%%%%%%%%%%%%

\section{Introduction}

The Skyrme model was originally made as a toy model for baryons where
the baryon is made of a soliton in a theory of pions with
higher-derivative terms \cite{Skyrme:1962vh,Skyrme:1961vq}.
The theory was taken much more serious after Witten showed that the
soliton, called the Skyrmion, is exactly the baryon in the large-$N_c$
limit of low energy QCD \cite{Witten:1983tw,Witten:1983tx}.
Although quite a few phenomenologically appealing results have been
achieved in the framework of the Skyrme model, see
e.g.~\cite{Battye:2009ad,Lau:2014baa,Halcrow:2014dsa,Karliner:2015qoa,Haberichter:2015qca}, 
a withstanding problem is that the binding energies are typically
about an order of magnitude too large, compared to experimental data. 
This motivated a large body of work attempting at lowering the binding
energies in Skyrme-like models. One direction is based on a
self-dual Yang-Mills theory in five dimensions, dimensionally reduced
to four dimensions and in turn giving rise to an infinite tower of
vector mesons \cite{Sutcliffe:2010et,Sutcliffe:2011ig}.
This leads one to search for a theory where the soliton -- the
Skyrmion -- is either a BPS state or saturates a BPS-like energy
bound. The original Skyrme model has an energy bound discovered long
ago by Faddeev \cite{Faddeev:1976pg}. Sometimes this bound is called a
Bogomol'nyi bound \cite{Bogomolny:1975de}, which may be misleading because the
target space of the Skyrme model is $S^3$, which is not K\"ahler and
hence cannot be supersymmetrized. 
The mentioned energy bound is, however, not
saturable unless the space is isometric to a
3-sphere \cite{Manton:1986pz}. 
A different model was constructed later, which is by now called the 
BPS Skyrme model as it has a saturable energy
bound \cite{Adam:2010fg,Adam:2010ds}.
Supersymmetrizing the BPS Skyrme model was attempted
in \cite{Queiruga:2015xka}, which however is not possible due to the
fact that its target space is not K\"ahler \cite{Zumino:1979et}. 

Supersymmetrizing the Skyrme model was attempted early on, in the
literature \cite{Bergshoeff:1984wb}. Several problems were encountered
along the way. The first problem of $S^3$ not being K\"ahler was
remedied by switching the target space to $\mathbb{C}P^1$, which
however is the target space of a baby Skyrme model rather than of the 
Skyrme model.  
The next
problem is that unless care is taken in constructing the model, one
will encounter the auxiliary field problem; i.e.~the auxiliary field,
$F$, in the chiral multiplet will have derivatives acting on it and in
turn making it a dynamical field\footnote{For a recent work in this
direction, however, see \cite{Fujimori:2016udq}. }.
Finally, even circumventing the auxiliary field problem, the
supersymmetrized Skyrme-like term turned out to have four time
derivatives \cite{Bergshoeff:1984wb,Freyhult:2003zb},
making a Hamiltonian formulation impossible.
The later attempt at supersymmetrizing the baby-Skyrme model also
arrived at the same type of Lagrangian with four time
derivatives \cite{Freyhult:2003zb}.

Thirty years after the problem was laid out in the seminal
paper \cite{Bergshoeff:1984wb} not much progress on constructing the
Skyrme term was made, until a systematic investigation of the
supersymmetric four-derivative term without the auxiliary field
problem \cite{Khoury:2010gb,Adam:2013awa,Nitta:2014pwa,Nitta:2015uba}
lead to the idea that a non-Abelian nontrivial solution to the
non-dynamic auxiliary field equation could produce a supersymmetric
Skyrme term. This was carried out in \cite{Gudnason:2015ryh} and
indeed the nontrivial solution to the auxiliary field equation gave a
term whose Nambu-Goldstone (NG) submanifold is exactly the Skyrme
term. 
The construction is based on complexifying SU(2) to
SL(2,$\mathbb{C}$), which in turn gives rise to three new bosonic
degrees of freedom called quasi-NG bosons 
\cite{Nitta:2014fca}; this complexification is inevitable due to
nonlinear realization theory. 
Once the quasi-NG bosons are turned on, the supersymmetric Skyrme term
is more complicated and does possess four time derivatives, again not
allowing for a Hamiltonian formulation.
The restriction to the NG submanifold, however, eliminates the four
time derivatives and as mentioned above, yields exactly the Skyrme
term. 
A twist compared to the non-supersymmetric Skyrme model, is that if
one tries to turn on a standard kinetic (Dirichlet) term, the
nontrivial solution to the auxiliary field equation will simply
eliminate it, leaving just a potential term for the quasi-NG bosons.
Introducing a superpotential has not been carried out yet. 

It requires an attractive term in the Lagrangian in order for
a Skyrmion to be stabilized -- i.e.~a nontrivial solution to the virial
equation due to Derrick's theorem. 
In the conventional Skyrme model, the kinetic term and 
the Skyrme term is balanced.
In this paper we will induce a
kinetic term in the theory by performing Scherk-Schwarz
(SS) \cite{Scherk:1978ta,Scherk:1979zr}
dimensional reduction (DR) in order to construct a Skyrmion,
although our Skyrmion is rather an instanton in three Euclidean
dimensions.
Although the original formulation breaks supersymmetry due to twisting
of the $R$-symmetry, supersymmetry preserving dimensional reductions
are possible, see e.g.~\cite{Eto:2006pg}. 

We derive 3-dimensional and 2-dimensional Lagrangians by supersymmetry
preserving SS dimensional reductions and in turn construct solitons in
all dimensions from two through four. The first soliton is the
instanton found by Speight \cite{Speight:2007ax} in the so-called pure
Skyrme model -- which is exactly the NG part of the supersymmetric
Skyrme model. 
Next, we construct a 3-dimensional Skyrmion-instanton 
in the once SS reduced theory.
Finally, we create a vortex-instanton in the twice SS anisotropically
reduced Euclidean two-dimensional theory.
All of these instantons are non-BPS states, breaking all supersymmetries.

The plan of the paper is as follows. We start by reviewing the
supersymmetric Skyrme model in Sec.~\ref{sec:model}.
In Sec.~\ref{sec:SSDR} we construct the SS dimensionally reduced
Lagrangians that we will use to construct the lower-dimensional
solitons in Sec.~\ref{sec:solitons}. Finally, we conclude with a
discussion in Sec.~\ref{sec:discussion}. 
The appendix discusses the BPS property of the solitons.

\section{The supersymmetric Skyrme model}\label{sec:model}

We will begin by reviewing the supersymmetric Skyrme model, found in
\cite{Gudnason:2015ryh}. 
The construction of fourth-order derivative terms in supersymmetry -- 
without the auxiliary field problem -- is based on the Lagrangian
\beq
\mathcal{L} = \int d^4 \theta \; K(\Phi,\Phi^{\dagger}) + 
\frac{1}{16}\int d^4\theta \; \Lambda_{AB \bar{C}\bar{D}}(\Phi,\Phi^\dag)
D^\alpha\Phi^A D_\alpha\Phi^B
\bar{D}_{\dot{\alpha}}\Phi^{\bar{C}\dag}\bar{D}^{\dot{\alpha}}\Phi^{\bar{D}\dag},
\label{eq:superfield_Lagrangian}
\eeq
where $K(\Phi,\Phi^{\dagger})$ is a K\"ahler potential and 
$\Lambda_{AB \bar{C} \bar{D}} (\Phi, \Phi^{\dagger})$ is a K\"ahler
tensor with the indices $A,B$ and $\bar{C},\bar{D}$ symmetrized
pairwise. 
The chiral superfields $\Phi^A$ are then combined with a nonlinear
sigma model field 
\beq
M = \exp(i\Phi^A t^A) \in G^{\mathbb{C}}/\hat{H},
\label{eq:NLSM_field}
\eeq
taking value in the coset relevant for chiral symmetry breaking:
\beq
G^{\mathbb{C}}/\hat{H} \simeq \mathrm{SU}(N)^{\mathbb{C}}
= G^{\mathbb{C}}/H^{\mathbb{C}} \simeq \mathrm{SL}(N,\mathbb{C})
\simeq T^*\mathrm{SU}(N).
\eeq
Here, $\hat{H}$ is the complex isotropy group and not necessarily equal to
$H^{\mathbb{C}}$ but can be larger in
general \cite{Bando:1983ab,Bando:1984cc,Bando:1984fn}.  

The superfields $\Phi^A$ are composed of NG bosons $\pi^A$, quasi-NG
bosons $\sigma^A$, quasi-NG fermions $\psi^A$ and auxiliary fields
$F^A$ as
\beq
\Phi^A(y,\theta) = \pi^A(y) + i\sigma^A(y) + \theta\psi^A(y)
+ \theta^2 F^A(y).
\eeq
Our case of chiral symmetry breaking falls
into the class of maximally realized supersymmetrizations (and
therefore $\hat{H}=H^{\mathbb{C}}$) which means that the number of
quasi-NG bosons is equal to the number of NG bosons \cite{Kotcheff:1988ji,Nitta:2014fca}.
The K\"ahler potential, $K(\Phi,\Phi^\dag)$, used for constructing the
supersymmetric Skyrme model \cite{Gudnason:2015ryh} is
\beq
K = f_\pi^2 \Tr M M^\dag,
\eeq
and the $(2,2)$ K\"ahler tensor,
$\Lambda_{AB\bar{C}\bar{D}}(\Phi,\Phi^\dag)$, is implicitly defined by  
\begin{align}
&\int d^4\theta\; \Lambda_{AB\bar{C}\bar{D}}(\Phi,\Phi^\dag)
D^\alpha \Phi^A
D_\alpha \Phi^B \bar{D}_{\dot{\alpha}}\Phi^{\bar{C}\dag} 
\bar{D}^{\dot{\alpha}}\Phi^{\bar{D}\dag} \non
&= \int d^4\theta\; \Lambda(M,M^\dag)\Tr\left[
D^\alpha M \bar{D}_{\dot{\alpha}} M^\dag D_\alpha
M \bar{D}^{\dot{\alpha}} M^\dag\right].
\end{align}
Then the bosonic part of the Lagrangian is
\begin{align}
\mathcal{L}_b^{(4)} &= \ f_{\pi}^2 \Tr
\left[-M_{\mu} M^{\mu} + F^{\dagger} F\right] \non
&\phantom{=\ }
+ \Lambda(M,M^\dag)\Tr\left[
M_\mu^\dag M_\nu M^{\mu\dag} M^\nu
+ (F^\dag F)^2
- M_\mu^\dag M^\mu F^\dag F
- M_\mu^\dag M^\mu F F^\dag\right],
\label{eq:L4}
\end{align}
where we have introduced the short notation $M_\mu\equiv\p_\mu M$
and $f_{\pi}$ is the pion decay constant.
The first term is the ordinary kinetic (Dirichlet) term with two time
derivatives while the second term is a higher derivative correction. 
The K\"ahler tensor $\Lambda_{AB\bar{C}\bar{D}}$ in
\eqref{eq:superfield_Lagrangian}
is determined by the $G$-invariant function $\Lambda (M,M^{\dagger})$
through the relation \eqref{eq:NLSM_field}.
The equations of motion for the auxiliary fields are
\begin{align}
f_{\pi}^2 \Lambda^{-1} F +
2 F F^{\dagger} F - F M^{\dagger}_{\mu} M^{\mu} - M_{\mu} M^{\dagger
 \mu} F &= 0, \non
f_{\pi}^2 \Lambda^{-1} F^{\dagger} + 
2 F^{\dagger} F F^{\dagger} - M_{\mu}^{\dagger} M^{\mu} F^{\dagger} -
 F^{\dagger} M_{\mu} M^{\dagger \mu} &= 0.
\label{eq:auxiliary_eom}
\end{align}
The Lagrangian \eqref{eq:L4} avoids the auxiliary field problem and
hence the auxiliary field equation is algebraic; it is, nevertheless,
a nontrivial matrix equation.

Two consistent possibilities of solutions to the equations
\eqref{eq:auxiliary_eom} arise if we do not introduce a superpotential.
The first corresponds to the trivial solution $F=0$ which is called the canonical
branch.
The on-shell Lagrangian on the canonical branch is obtained
straightforwardly from 
eq.~\eqref{eq:L4}:
\beq
\mathcal{L}_b^{(4)} = - f_{\pi}^2 \Tr M^{\dagger}_{\mu} M^{\mu} + 
\Lambda(M,M^\dag)\Tr\left[
M_\mu^\dag M_\nu M^{\mu\dag} M^\nu\right].
\eeq
However, this term does not reduce to the Skyrme term when the
quasi-NG fields are set to zero and it also contains four
time-derivatives.

The second is the non-canonical branch associated with the non-trivial solutions
$F\neq 0$.
The non-canonical branch for the theory \eqref{eq:L4} without a
superpotential was found explicitly in \cite{Gudnason:2015ryh}
for the SU(2) case
\begin{align}
&\mathcal{L}_{b}^{(4)} = 
\frac{\Lambda(M,M^\dag)}{2} \bigg\{\Tr\left[ 
  2M_\mu^\dag M_\nu M^{\mu\dag} M^\nu
  -\frac{1}{2}M_\mu^\dag M^\mu M_\nu^\dag M^\nu
  -\frac{1}{2}M_\mu M^{\mu\dag} M_\nu M^{\nu\dag}\right]
  \label{eq:Lexplicit}\\
  &-\frac{1}{2}\left(\Tr[M_\mu M^{\mu\dag}]\right)^2
- \Tr\left[\frac{f^4_{\pi}}{2 \Lambda^2 (M,M^{\dagger})} \mathbf{1}_2\right]
\non
  &\mp\sqrt{\left(
    \Tr[M_\mu^\dag M^\mu M_\nu^\dag M^\nu]
    -\frac{1}{2}\left(\Tr[M_\mu M^{\mu\dag}]\right)^2\right)
    \left(
    \Tr[M_\mu M^{\mu\dag} M_\nu M^{\nu\dag}]
    -\frac{1}{2}\left(\Tr[M_\mu M^{\mu\dag}]\right)^2\right)}
    \bigg\}. \nonumber
\end{align}
We note that the ordinary second-order kinetic term is canceled on the
non-canonical branch.
Although the term with four time derivatives does not cancel in
general when the quasi-NG fields are turned on, the above Lagrangian
simplifies exactly to the Skyrme term when they are turned off
\beq
\left.\mathcal{L}_{b}^{(4)}\right|_{M=U} =
\Lambda \Tr\left[
  U_\mu^\dag U_\nu U^{\mu\dag} U^\nu
  - U_\mu^\dag U^\mu U_\nu^\dag U^\nu
\right]
- \Tr 
\left[
\frac{f_{\pi}^4}{4 \Lambda} \mathbf{1}_2
\right]. \label{eq:LSkyrme}
\eeq
The prefactor $\Lambda(M,M^\dag)$ is a function of $G$-invariants and
thus when restricting to the NG submanifold,
$\Lambda(U U^\dag=\mathbf{1}_2)=\Lambda$ becomes a constant.
The effect of adding the kinetic (Dirichlet) term thus only had the
effect of inducing the potential (the last term in the above
Lagrangian).
If we consider a $G$-invariant theory, $\Lambda$ must be a constant
and thus the potential is just a c-number that we can ignore.

The upshot is thus that the K\"ahler potential cannot induce a kinetic 
term or a potential if $G$-invariance is preserved.
In the next section, we will perform Scherk-Schwarz dimensional
reductions in order to induce a kinetic term.

\section{Scherk-Schwarz dimensional reductions}\label{sec:SSDR}

Our starting point will be the supersymmetric Skyrme
model \eqref{eq:Lexplicit} in 4-dimensional Euclidean space. 
We will keep the $G$-invariance intact
throughout this paper, so the addition of the kinetic term will only
induce a constant and hence not affect the equations of motion.
Therefore we will simply work with only the fourth-order derivative
term and ignore the latter constant by setting $f_{\pi} = 0$. % why do
                                % we have to set f_\pi = 0 ?
For simplicity, we will restrict to the NG submanifold before
performing Scherk-Schwarz (SS) dimensional reduction (DR) and it will
prove convenient to change notation to a 4-vector
$\mathbf{n}=\{n^1,n^2,n^3,n^4\}$ as 
\beq
U = \mathbf{1}_2 n^4 + i n^a \tau^a,
\eeq
where $\tau^a$ are the Pauli matrices and the fields satisfy
$\mathbf{n}\cdot \mathbf{n} = 1$.
Using this notation, the NG restricted Skyrme model on Euclidean
four-space, $\mathbb{R}^4$, reads
\beq
\mathcal{L}_{4d,b}^{(4)} =
\frac{1}{4}(\mathbf{n}_\mu\cdot\mathbf{n}_\mu)^2
-\frac{1}{4}(\mathbf{n}_\mu\cdot\mathbf{n}_\nu)^2,
\label{eq:Ln4d}
\eeq
where we have defined $\mathbf{n}_{\mu} = \partial_{\mu} \mathbf{n}$ and 
lowered all the indices since the Euclidean metric is
just the identity matrix. We have also set $\Lambda=1/16$ for
convenience.
The Lagrangian \eqref{eq:Ln4d} admits a symmetry under the
transformation $\mathbf{n}'= O \mathbf{n}, \ O \in \mathrm{SO}(4)$
which will be utilized for the SS reduction.

\subsection{Three-dimensional model}

We are now ready to perform the first SSDR by compactifying the fourth
coordinate. We will use the coordinates $x^\mu$ with $\mu=1,2,3,4$ 
where we have Wick-rotated the time coordinate, and 
SS dimensional reduction along $x^4\sim x^4+2\pi R_4$ is carried out
as follows
\beq
\mathbf{n}(x^\mu) = O(x^4)\mathbf{N}(x^a),
\eeq
where $a=1,2,3$ runs over the non-compactified dimensions and the
matrix
\beq
O(x^4)=-O(x^4+2\pi R_4)\equiv
\begin{pmatrix}
\cos\frac{\mathsf{m}_{4,1}x^4}{2R_4} & -\sin\frac{\mathsf{m}_{4,1}x^4}{2R_4} & 0 & 0\\
\sin\frac{\mathsf{m}_{4,1}x^4}{2R_4} & \cos\frac{\mathsf{m}_{4,1}x^4}{2R_4} & 0 & 0\\
0 & 0 & \cos\frac{\mathsf{m}_{4,2}x^4}{2R_4} & -\sin\frac{\mathsf{m}_{4,2}x^4}{2R_4}\\
0 & 0 & \sin\frac{\mathsf{m}_{4,2}x^4}{2R_4} & \cos\frac{\mathsf{m}_{4,2}x^4}{2R_4}
\end{pmatrix},
\eeq
where $\mathsf{m}_{4,1}\in\mathbb{Z}_{\neq 0}$ and
$\mathsf{m}_{4,2}\in\mathbb{Z}_{\neq 0}$ are two nonzero Kaluza-Klein
(KK) integers describing towers of higher-momentum states along the
compactified circle.
Notice that we have applied twisted boundary conditions (TBC) such
that
\beq
\mathbf{n}(x^4+2\pi R_4) =
\epsilon(\mathsf{m}_{4,1},\mathsf{m}_{4,2})\mathbf{n}(x^4),
\eeq
with
\beq
\epsilon(\mathsf{m}_{4,1},\mathsf{m}_{4,2})=
\begin{pmatrix}
(-1)^{\mathsf{m}_{4,1}} \\
&(-1)^{\mathsf{m}_{4,1}} \\
&&(-1)^{\mathsf{m}_{4,2}} \\
&&&(-1)^{\mathsf{m}_{4,2}}
\end{pmatrix},
\label{eq:epsilon_matrix}
\eeq
which is just an SO(4) transformation of the fields.

After the dust settles, we obtain
\begin{align}
\mathcal{L}_{3d,b}^{(4)} &=
\frac{\pi R_4}{2}\left[
(\mathbf{N}_a\cdot\mathbf{N}_a)^2
-(\mathbf{N}_a\cdot\mathbf{N}_b)^2\right] \non
&\phantom{=\ }
+\frac{\pi}{4R_4}\left[\mathsf{m}_{4,1}^2
+ (\mathsf{m}_{4,2}^2-\mathsf{m}_{4,1}^2)\{(N^3)^2+(N^4)^2\}\right]
\mathbf{N}_a\cdot\mathbf{N}_a \non
&\phantom{=\ }
-\frac{\pi}{4R_4}\left[
\mathsf{m}_{4,1}(N^1N_a^2 - N^2N_a^1)
+ \mathsf{m}_{4,2}(N^3N_a^4 - N^4N_a^3)\right]^2, \label{eq:L3m1m2}
\end{align}
where we again have defined the notation
$\mathbf{N}_a = \partial_a \mathbf{N}$
and used the relation $\mathbf{N}\cdot\mathbf{N}=1$. 
If we set the two integers $\mathsf{m}_{4,1}$ and $\mathsf{m}_{4,2}$
equal to each other and to $\mathsf{m}_4\in\mathbb{Z}_{\neq 0}$, the SS
dimensionally reduced Lagrangian simplifies to
\begin{align}
\mathcal{L}_{3d,b}^{(4)} &=
\frac{\pi R_4}{2}\left[
(\mathbf{N}_a\cdot\mathbf{N}_a)^2
-(\mathbf{N}_a\cdot\mathbf{N}_b)^2\right]
+\frac{\mathsf{m}_4^2\pi}{4R_4}
\mathbf{N}_a\cdot\mathbf{N}_a \non
&\phantom{=\ }
-\frac{\mathsf{m}_4^2\pi}{4R_4}\left[
N^1N_a^2 - N^2N_a^1 + N^3N_a^4 - N^4N_a^3\right]^2.
\label{eq:L3}
\end{align}
Then the last term in the first line is a kinetic term with a
prefactor of the KK mass in the 3-dimensional Euclidean theory. 
We stress that the SS reduction of the fourth-order derivative term of
the BPS Skyrme model produces the usual (second-order derivatives
term) kinetic term. 
This is in contradistinction to the ordinary case where the potential
term appears by SS reduction from the usual kinetic term.

Let us note that the lowest energy state comes from the lowest KK mode
and thus the compactified momenta correspond to the integers
$\mathsf{m}_{4,1}=\pm 1$ and $\mathsf{m}_{4,2}=\pm 1$.
It is clear from the SS reduced Lagrangian \eqref{eq:L3} that the
overall sign of the two integers is physically unobservable.
One may naively think that the relative sign could matter, but
renaming the two fields $\{N^3,N^4\}\to\{N^4,N^3\}$ compensates a
relative minus sign.

A further remark about the KK momenta is in store. Because
$\pi_1(\mathrm{SU}(2))$ is trivial, higher even momentum numbers
$\mathsf{m}_4$ may be metastable or unstable and could decay to
$\mathsf{m}_{4}=0$ while for odd integers, they may decay to the
states with $\mathsf{m}_4=\pm 1$. This holds for both the KK
integers. The states with minimum energy are thus
$\mathsf{m}_4=-1,0,1$, where the $\mathsf{m}_4=0$ is distinguished
from $\mathsf{m}_4=\pm 1$ by the boundary conditions.
We will focus on the latter in this paper.

\subsection{Two-dimensional model}

Now we will perform another consecutive SS dimensional reduction, but
along $x^3\sim x^3+2\pi R_3$ as
\beq
\mathbf{N}(x^a) = \widetilde{O}(x^3)\mathbf{M}(x^i),
\eeq
where $i=1,2$ runs over the non-compactified dimensions and
\beq
\widetilde{O}(x^3) = -\widetilde{O}(x^3+2\pi R_3) \equiv
\begin{pmatrix}
\cos\frac{\mathsf{m}_{3,1}x^3}{2R_3} & -\sin\frac{\mathsf{m}_{3,1}x^3}{2R_3} & 0 & 0\\
\sin\frac{\mathsf{m}_{3,1}x^3}{2R_3} & \cos\frac{\mathsf{m}_{3,1}x^3}{2R_3} & 0 & 0\\
0 & 0 & \cos\frac{\mathsf{m}_{3,2}x^3}{2R_3} & -\sin\frac{\mathsf{m}_{3,2}x^3}{2R_3}\\
0 & 0 & \sin\frac{\mathsf{m}_{3,2}x^3}{2R_3} & \cos\frac{\mathsf{m}_{3,2}x^3}{2R_3}
\end{pmatrix},
\eeq
where $\mathsf{m}_{3,1}\in\mathbb{Z}_{\neq 0}$ is a nonzero  
integer while $\mathsf{m}_{3,2}\in\mathbb{Z}$ is an integer; we allow
it to be vanishing in order to get an anisotropic SS dimensional
reduction (this does not correspond to a compactified momentum not
being quantized, but merely formally to the option of making the last
two fields independent of the circle coordinate).
The TBC are then
$\mathbf{N}(x^3+2\pi R_3)=\epsilon(\mathsf{m}_{3,1},\mathsf{m}_{4,2})\mathbf{N}(x^3)$, with
$\epsilon$ again given by eq.~\eqref{eq:epsilon_matrix}. 

Starting now from the 3-dimensional Euclidean
Lagrangian \eqref{eq:L3m1m2}, we get
\begin{align}
\mathcal{L}_{2d,b}^{(4)} &=
\pi^2 R_3 R_4 \left[
(\mathbf{M}_i\cdot\mathbf{M}_i)^2
-(\mathbf{M}_i\cdot\mathbf{M}_j)^2\right] \non
&\phantom{=\ }
+\frac{\pi^2}{2}\bigg[
  \mathsf{m}_{4,1}^2\frac{R_3}{R_4}
  +\mathsf{m}_{3,1}^2\frac{R_4}{R_3} \non
&\phantom{=+2\pi^2\bigg[\ }
  +\left\{(\mathsf{m}_{4,2}^2-\mathsf{m}_{4,1}^2)\frac{R_3}{R_4}
    +(\mathsf{m}_{3,2}^2-\mathsf{m}_{3,1}^2)\frac{R_4}{R_3}\right\}
    \left[(M^3)^2+(M^4)^2\right]\bigg]
  \mathbf{M}_i\cdot\mathbf{M}_i \non
&\phantom{=\ }
-\frac{\pi^2}{2}\left(\mathsf{m}_{4,1}^2\frac{R_3}{R_4}
  +\mathsf{m}_{3,1}^2\frac{R_4}{R_3}\right)
  \left(M^1 M_i^2 - M^2 M_i^1\right)^2 \non
&\phantom{=\ }
-\frac{\pi^2}{2}\left(\mathsf{m}_{4,2}^2\frac{R_3}{R_4}
  +\mathsf{m}_{3,2}^2\frac{R_4}{R_3}\right)
  \left(M^3 M_i^4 - M^4 M_i^3\right)^2 \non
&\phantom{=\ }
-\pi^2\left(\mathsf{m}_{4,1}\mathsf{m}_{4,2}\frac{R_3}{R_4}
  +\mathsf{m}_{3,1}\mathsf{m}_{3,2}\frac{R_4}{R_3}\right)
  \left(M^1 M_i^2 - M^2 M_i^1\right)\left(M^3 M_i^4 - M^4 M_i^3\right) \non
&\phantom{=\ }
+\frac{\pi^2}{2R_3 R_4}\left(
  \mathsf{m}_{4,1}^2\mathsf{m}_{3,2}^2
  +\mathsf{m}_{4,2}^2\mathsf{m}_{3,1}^2
  -2\mathsf{m}_{3,1}\mathsf{m}_{3,2}\mathsf{m}_{4,1}\mathsf{m}_{4,2}\right) \non
&\phantom{=+\frac{2\pi^2}{R_3 R_4}\ } \times
\left[(M^1)^2 + (M^2)^2\right]\left[(M^3)^2 + (M^4)^2\right],
\label{eq:L2m1m2}
\end{align}
where $\mathbf{M}_i = \partial_i \mathbf{M}$.
In the case the two momenta on each compactified circle are equal,
viz.~when $\mathsf{m}_{3,1}=\mathsf{m}_{3,2}=\mathsf{m}_3$ and
$\mathsf{m}_{4,1}=\mathsf{m}_{4,2}=\mathsf{m}_4$, a great
simplification occurs
\begin{align}
\mathcal{L}_{2d,b}^{(4)} &=
\pi^2 R_3 R_4 \left[
(\mathbf{M}_i\cdot\mathbf{M}_i)^2
-(\mathbf{M}_i\cdot\mathbf{M}_j)^2\right]
+\frac{\pi^2}{2}\left[
  \mathsf{m}_{4}^2\frac{R_3}{R_4}
  +\mathsf{m}_{3}^2\frac{R_4}{R_3}\right]
  \mathbf{M}_i\cdot\mathbf{M}_i \non
&\phantom{=\ }
-\frac{\pi^2}{2}\left(\mathsf{m}_{4}^2\frac{R_3}{R_4}
  +\mathsf{m}_{3}^2\frac{R_4}{R_3}\right)
  \left(M^1 M_i^2 - M^2 M_i^1 + M^3 M_i^4 - M^4 M_i^3\right)^2.
\end{align}
However, we can see that this simplification also eliminates the
potential, i.e.~only derivative terms remain. 

In order to get a relatively simple Lagrangian with a potential, let
us consider the case where the first compactification has equal
momenta ($\mathsf{m}_{4,1}=\mathsf{m}_{4,2}=\mathsf{m}_4$) while the
second compactification has one nonvanishing momentum
$\mathsf{m}_{3,1}=\mathsf{m}_3\neq 0$ and one vanishing
$\mathsf{m}_{3,2}=0$. This will simplify the
Lagrangian \eqref{eq:L2m1m2} to
\begin{align}
\mathcal{L}_{2d,b}^{(4)} &=
\pi^2 R_3 R_4 \left[
(\mathbf{M}_i\cdot\mathbf{M}_i)^2
-(\mathbf{M}_i\cdot\mathbf{M}_j)^2\right] \non
&\phantom{=\ }
+\frac{\pi^2}{2}\bigg[
  \mathsf{m}_{4}^2\frac{R_3}{R_4}
  +\mathsf{m}_{3}^2\frac{R_4}{R_3}
    \left[(M^1)^2+(M^2)^2\right]\bigg]
  \mathbf{M}_i\cdot\mathbf{M}_i \non
&\phantom{=\ }
-\frac{\pi^2}{2}\mathsf{m}_{4}^2\frac{R_3}{R_4}
  \left(M^1 M_i^2 - M^2 M_i^1 + M^3 M_i^4 - M^4 M_i^3\right)^2 \non
&\phantom{=\ }
-\frac{\pi^2}{2}\mathsf{m}_{3}^2\frac{R_4}{R_3}
  \left(M^1 M_i^2 - M^2 M_i^1\right)^2 \non
&\phantom{=\ }
+\frac{\pi^2}{2R_3 R_4}\mathsf{m}_{4}^2\mathsf{m}_{3}^2
  \left[(M^1)^2 + (M^2)^2\right]\left[(M^3)^2 + (M^4)^2\right].
\label{eq:L2}
\end{align}
We note again that the lowest energy states correspond to the lowest
KK modes, being $\mathsf{m}_4=\pm 1$ and $\mathsf{m}_3=\pm 1$.
The two signs are obviously not observable in the above
Lagrangian \eqref{eq:L2}.
One may however ask whether what consequences a relative sign between
the two momenta on the first compactified circle
($x^4\sim x^4+2\pi R_4$) may yield. Again it simply amounts to a sign
in front of the last two terms (in the parenthesis) on the third line
in eq.~\eqref{eq:L2}, which again can easily be compensated by
renaming the two fields $\{M^3,M^4\}\to\{M^4,M^3\}$.

In the next section we will construct Euclidean solitons in the
above Lagrangians.

\section{Euclidean solitons or instantons}\label{sec:solitons}

In this section we will consider Euclidean solitons in the
supersymmetric Skyrme model and its derivatives coming from SS
dimensional reduction. 

\subsection{4d pure Skyrme instanton}

The first and simplest case is to consider a Euclidean soliton
directly in the 4-dimensional theory \eqref{eq:Ln4d}, since the
action is classically conformal in said number of dimensions.
It is therefore an instanton-like soliton, first constructed in
\cite{Speight:2007ax}.
Note that the pure Skyrme model \cite{Speight:2007ax} and the bosonic
sector of the NG restricted submanifold of the supersymmetric Skyrme
model are identical. 
Hence the solution is directly applicable and here we will just make a 
swift review of the pure Skyrme-instanton.
Let us start with the Lagrangian density \eqref{eq:LSkyrme} and use
the Ansatz for the Skyrme field
\beq
U = q \eta q^{-1},
\eeq
where $q$ is the identity map from $S^3\to\mathrm{SU}(2)$ as
\beq
q = \hat{x}^4\mathbf{1}_2 + i\tau^a\hat{x}^a,
\eeq
where $\hat{x}^\mu\equiv x^\mu/r$ and the radius of the 3-sphere is
$r=\sqrt{x^\mu x^\mu}$ and finally
\beq
\eta = \eta_0(r)\mathbf{1}_2 + i\tau^3\eta_3(r),
\eeq
where $\eta$ is a curve in SU(2) obeying the constraint
$\eta_0^2+\eta_3^2=1$.
Thus the Euclidean action can be written as
\beq
S_E = 16\pi^2\int ds \left[
\frac{1}{2}\eta_0^{\prime 2}(s) + \left[1-\eta_0^2(s)\right]^2\right],
\label{eq:SESkyrmeinstanton}
\eeq
where we have introduced $s=\log r$.
The Skyrme-instanton solution is
\beq
\eta_0(s) = \tanh \sqrt{2}(s-s_0),
\label{eq:skyrme-instanton}
\eeq
which solves both the second-order equation of motion derived from
the Euclidean action \eqref{eq:SESkyrmeinstanton} and the Bogomol'nyi
equation
\beq
\eta_0'(s) = \sqrt{2}\left[1-\eta_0^2(s)\right].
\eeq
The Euclidean action associated with the Skyrme-instanton solution
\eqref{eq:skyrme-instanton} is
\beq
S_E = 16\sqrt{2}\pi^2\int ds\; \eta_0'(s)\left[1 - \eta_0^2(s)\right]
= 16\sqrt{2}\pi^2 \int d\eta_0 \left[1-\eta_0^2\right]
= \frac{64\sqrt{2}\pi^2}{3},
\eeq
where we have used the boundary conditions for the instanton solution:
$\eta_0(-\infty)=-1$ and $\eta_0(\infty)=1$.
The topological charge of the instanton is, however,
$\mathbb{Z}_2$ \cite{Speight:2007ax}; to see this requires a
suspension of the Hopf map to get to a nontrivial $\pi_4(S^3)$, see
also e.g.~\cite{Nakahara:1986kn}. 

\begin{figure}[!htp]
\begin{center}
\includegraphics[width=0.5\linewidth]{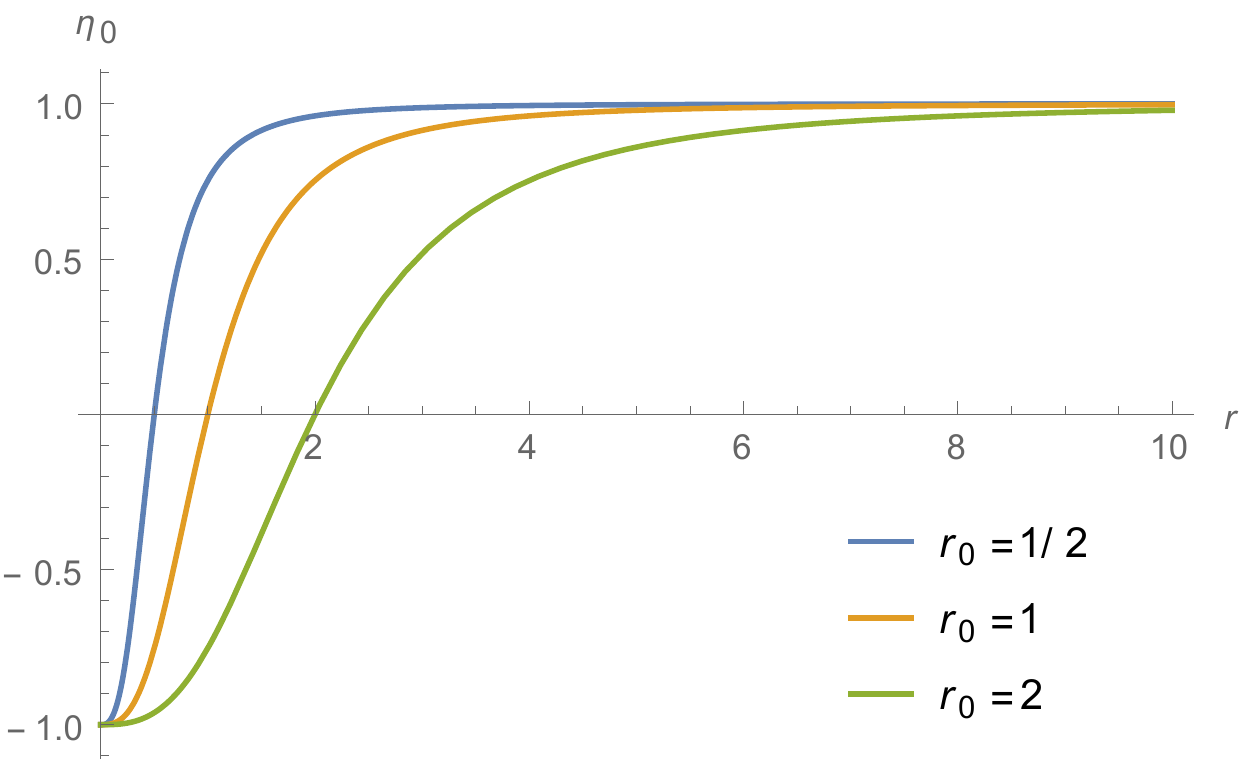}
\caption{Skyrme-instanton solutions for various instanton sizes
$r_0=e^{s_0}=1/2,1,2$. }
\label{fig:4dSkyrme-instanton}
\end{center}
\end{figure}

In Fig.~\ref{fig:4dSkyrme-instanton} we show 4d Skyrme-instanton
solutions for various instanton sizes. Notice that $\eta_0(r_0)=0$,
i.e.~the curve in SU(2) passes zero exactly at the values of
$r_0=e^{s_0}$ which we called the instanton size.

In the following, we look for instantons in three- and two-dimensional
models.

\subsection{3d Skyrmion-instanton}

Skyrmions in the pure Skyrme theory (without a kinetic or potential
term) are unstable against expanding themselves. Either a kinetic term
or a potential term is needed to stabilize the Skyrmions with a finite
size (and energy). 
Here, we have the kinetic term for the SS dimensionally reduced
theory and we call Skyrmions in 3-dimensional Euclidean theory 
Skyrmion-instantons. 
After performing a single SS dimensional reduction to 3-dimensional
Euclidean space, we have the Lagrangian density \eqref{eq:L3m1m2}.
For simplicity, we will construct a soliton in the isotropically
reduced theory, namely \eqref{eq:L3} which corresponds to the latter
Lagrangian with the two momenta on the circle set equal
$\mathsf{m}_{4,1}=\mathsf{m}_{4,2}$.

For convenience, we will rescale the lengths as
$x^a=\frac{2R_4}{\mathsf{m}_4}\tilde{x}^a$, where $\tilde{x}^a$ are 
dimensionless coordinates.
The Euclidean action thus reads
\begin{align}
S_E &= \pi\mathsf{m}_4 \int d^3\tilde{x}\; \mathcal{L}_E
\label{eq:SE-3d-Skyrme-instanton}\\
\mathcal{L}_E &=
\frac{1}{4}\left[
(\mathbf{N}_a\cdot\mathbf{N}_a)^2
-(\mathbf{N}_a\cdot\mathbf{N}_b)^2\right]
+\frac{1}{2}
\mathbf{N}_a\cdot\mathbf{N}_a 
-\frac{1}{2}\left[
N^1N_a^2 - N^2N_a^1 + N^3N_a^4 - N^4N_a^3\right]^2.
\nonumber
\end{align}
The equations of motion derived from the above action read
\begin{align}
&N_{aa}^\alpha
+(\mathbf{N}_b\cdot\mathbf{N}_b) N_{aa}^\alpha
+(\mathbf{N}_{ab}\cdot\mathbf{N}_b) N_a^\alpha
-(\mathbf{N}_a\cdot\mathbf{N}_b) N_{ab}^\alpha
-(\mathbf{N}_{bb}\cdot\mathbf{N}_a) N_a^\alpha \non
&+\left(N^1 N_{aa}^2 - N^2 N_{aa}^1 + N^3 N_{aa}^4 - N^4
N_{aa}^3\right)
  \left(N^2 \delta^{\alpha 1} - N^1 \delta^{\alpha 2} +
  N^4 \delta^{\alpha 3} - N^3 \delta^{\alpha 4}\right) \non
&+2\left(N^1 N_a^2 - N^2 N_a^1 + N^3 N_a^4 - N^4 N_a^3\right)
  \left(N_a^2 \delta^{\alpha 1} - N_a^1 \delta^{\alpha 2}
  + N_a^4 \delta^{\alpha 3} - N_a^3 \delta^{\alpha 4}\right) = 0,
\end{align}
where $\alpha=1,2,3,4$. 
We will call the soliton a 3d Skyrmion-instanton and it is very 
similar to a Skyrmion in the sense that it wraps a 3-sphere in the
target space and lives in 3-dimensional (Euclidean) configuration
space. 

\begin{figure}[!thp]
\begin{center}
\mbox{
\subfloat[]{\includegraphics[width=0.49\linewidth]{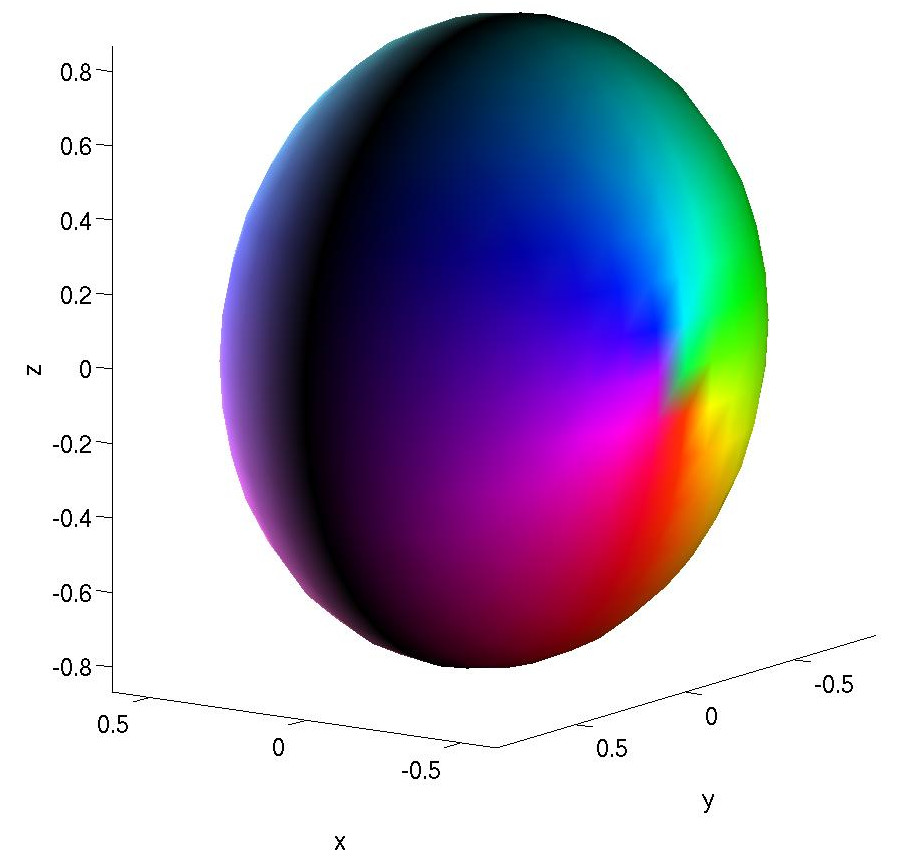}}
\subfloat[]{\includegraphics[width=0.49\linewidth]{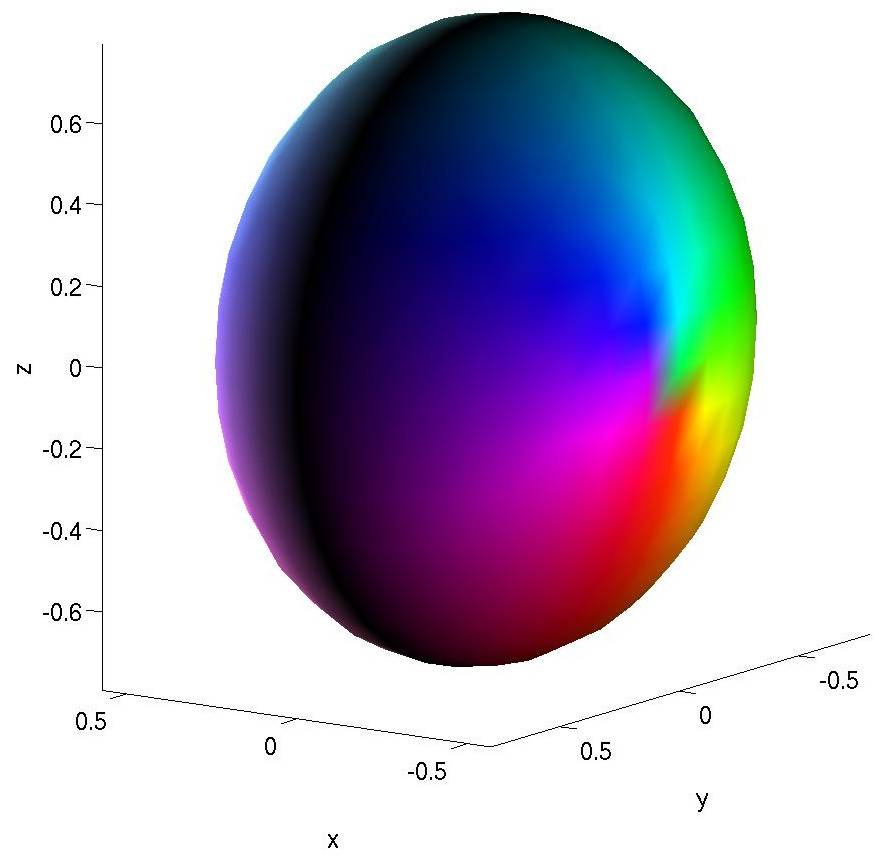}}}
\caption{3d-Skyrmion-instanton solution in the once SS reduced
3-Euclidean dimensional theory.
(a) shows the isosurface of the topological charge density and (b) 
the isosurface of the Euclidean Lagrangian density; both at their
respective half-maximum values.
The solution is a squashed sphere with squashing parameter measured to  
be $\sigma[\mathcal{B}]=0.9499$ and $\sigma[\mathcal{L}_E]=0.9308$;
the topological charge evaluated numerically as
$B^{\rm numerical}=0.9992$ and the Euclidean action is 
$S_E=\pi\mathsf{m}_4\times 55.91$.
The colors are described in the text.
}
\label{fig:3d-Skyrme-instanton}
\end{center}
\end{figure}

Since the last term in the action \eqref{eq:SE-3d-Skyrme-instanton}
breaks spherical symmetry we have not been able to reduce the equation 
of motion for a single 3d-Skyrmion-instanton to an ordinary
differential equation.
We therefore turn to numerical methods and solve the full partial
differential equations with the finite difference method in
conjunction with the relaxation method on an $81^3$ cubic lattice with
a fourth-order stencil.
We define the topological charge of the 3d-Skyrmion-instanton as 
\beq
B = -\frac{1}{2\pi^2} \int d^3\tilde{x}\; \mathcal{B}, \qquad
\mathcal{B} = \frac{1}{6}
\epsilon^{abc}\epsilon^{\alpha\beta\gamma\delta}
N_a^\alpha N_b^\beta N_c^\gamma N^\delta.
\eeq
The solution is shown as isosurfaces of the topological charge and
Euclidean Lagrangian densities at their respective half-maximum values
in Fig.~\ref{fig:3d-Skyrme-instanton}.
We have colored the figures using a normalized 3-vector
$\mathbf{v}\equiv (N^2,N^3,N^4)/\sqrt{(N^2)^2+(N^3)^2+(N^4)^2}$ and mapping
$v^3+iv^2=e^{i\theta_{\rm color}}$, where
$\theta_{\rm color}=0,\pi/3,2\pi/3$ corresponds to red, green and blue,
respectively. $v^1$ is then mapped to the lightness with $|v^1|=1$
being white and $v^1=0$ being black. 

As we can see from the figure, the solution is a squashed sphere.
In order to calculate the squashing of the solution, let us first 
define the size of the 3d-Skyrmion-instanton along the $x^a$ direction
for a fixed $a$ as 
\beq
\langle (x^a)^2[\mathcal{X}]\rangle =
\frac{\int d^3\tilde{x}\; (x^a)^2\mathcal{X}}
{\int d^3\tilde{x}\; \mathcal{X}},
\eeq
where the index $a$ is \emph{not} summed over and $\mathcal{X}$ is a
density.
The solution calculated in Fig.~\ref{fig:3d-Skyrme-instanton} is
rotated such that there is an axial symmetry in the $(x^1,x^3)$-plane
and thus we can define the squashing parameter as
\beq
\sigma[\mathcal{X}]\equiv
\sqrt{\frac{\langle (x^2)^2[\mathcal{X}]\rangle}{\langle
(x^1)^2[\mathcal{X}]\rangle}},
\eeq
where we will use the topological charge density
$\mathcal{X}=\mathcal{B}$ and the Euclidean Lagrangian density
$\mathcal{X}=\mathcal{L}_E$, respectively; the numerical calculation
gives $\sigma[\mathcal{B}]=0.9499$ and 
$\sigma[\mathcal{L}_E]=0.9308$.

\subsection{2d vortex-instanton}

We will now consider the case of two consecutive SS dimensional
reductions where the first one is isotropic and the second dimensional
reduction is anisotropic in the way that only the first two fields
depend nontrivially on the second circle coordinate.
We thus have the case of the first reduction with equal momenta
($\mathsf{m}_{4,1}=\mathsf{m}_{4,2}=\mathsf{m}_4$) and the second
reduction with only one momentum ($\mathsf{m}_{3,1}=\mathsf{m}_3$ and
formally $\mathsf{m}_{3,2}=0$).

It will again prove convenient to rescale the lengths as
$x^i=2\sqrt{\frac{R_3 R_4}{\mathsf{m}_3\mathsf{m}_4}}\tilde{x}^i$,
where $\tilde{x}^i$ are dimensionless coordinates. 
The anisotropic dimensional reduction induces a potential which we
need in order to construct a Euclidean vortex.
We will employ the appropriate Ansatz for the
vortex \cite{Gudnason:2014gla,Gudnason:2014hsa,Gudnason:2016yix} 
\beq
\mathbf{M} = \left\{
\cos f(r)\cos\alpha,\cos f(r)\sin\alpha,
\sin f(r)\cos\theta,\sin f(r)\sin\theta\right\},
\eeq
where $r e^{i\theta}=\tilde{x}^1+i\tilde{x}^2$ are the standard polar
coordinates in two dimensions and $\alpha$ is a U(1) modulus. 
Finally, the 2-dimensional Euclidean action for the vortex system is
found by plugging the above Ansatz into the Lagrangian \eqref{eq:L2}
and it reads 
\begin{align}
S_E &= \pi^3\mathsf{m}_3\mathsf{m}_4 \int dr\; r\mathcal{L}_E, \\
\mathcal{L}_E &=
\frac{1}{r^2}\sin^2(f) f_r^2
+\left(\frac{1}{\kappa} + \kappa\cos^2 f\right)
  \left(f_r^2 + \frac{1}{r^2}\sin^2 f\right)
-\frac{1}{\kappa r^2} \sin^4 f
+\cos^2 f\sin^2 f, \label{eq:L2dvortex}
\end{align}
where we have defined
\beq
\kappa \equiv \frac{\mathsf{m}_3 R_4}{\mathsf{m}_4 R_3}.
\eeq
The winding number is defined as
\beq
\mathcal{N} = \frac{1}{2\pi} \int dr\;
\left(M_r^3 M_\theta^4 - M_r^4 M_\theta^3\right)
= \frac{1}{2\pi}\int dr\; \sin(2f) f_r
= \frac{1}{2\pi}\int df\; \sin 2f = 1,
\eeq
where we have used the boundary conditions $f(0)=0$ and
$f(\infty)=\pi/2$ in the last equality. 

\begin{figure}[!thp]
\begin{center}
\mbox{
\subfloat[$f$]{\includegraphics[width=0.49\linewidth]{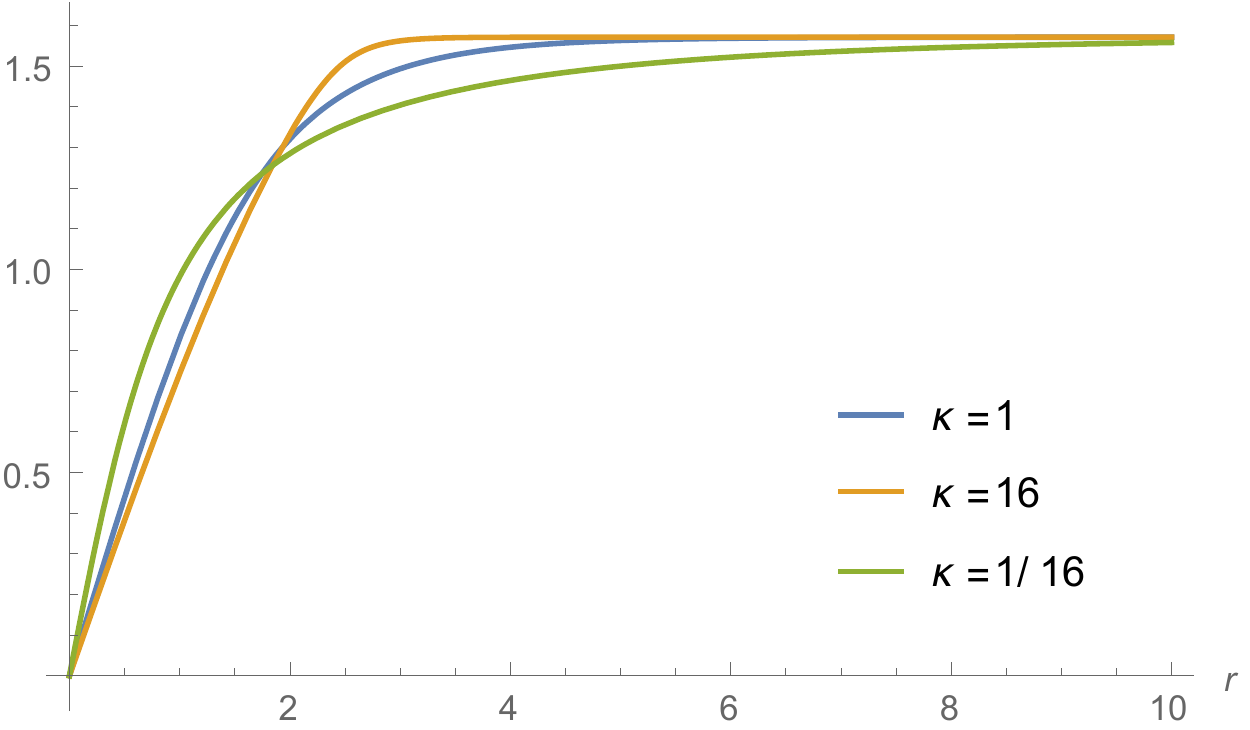}}
\subfloat[$\mathcal{L}_E$]{\includegraphics[width=0.49\linewidth]{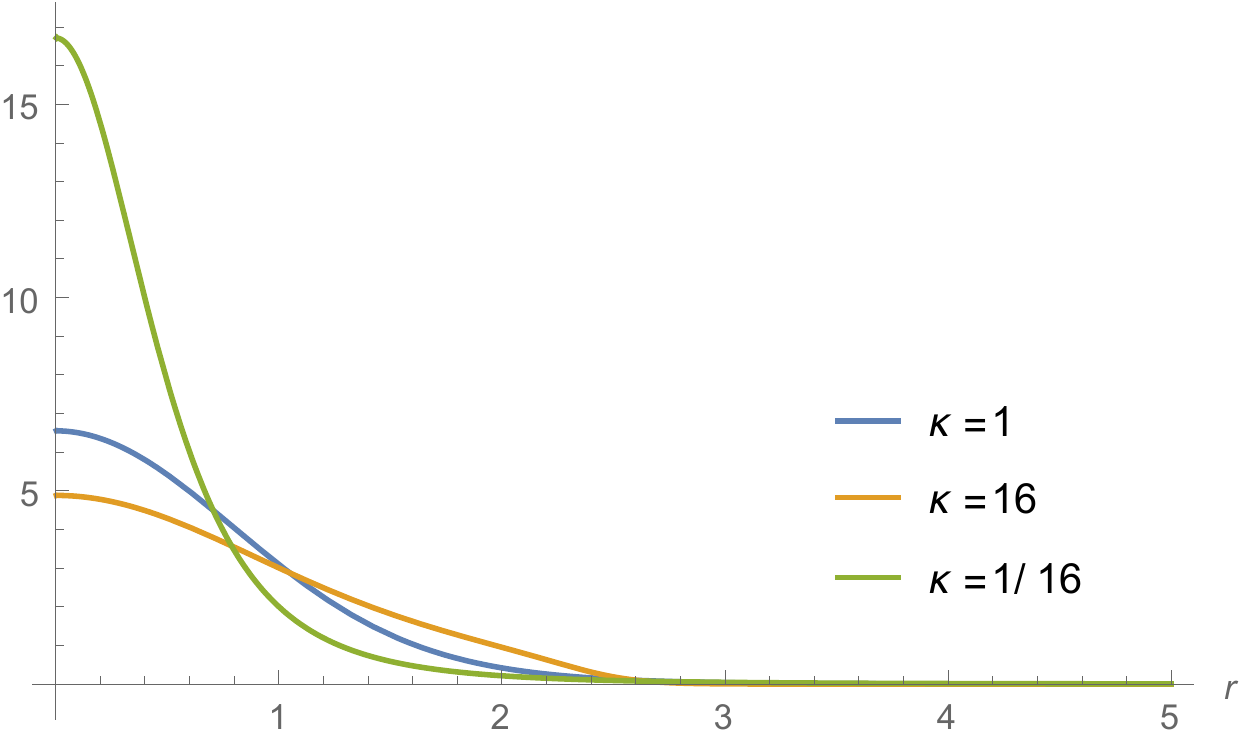}}}
\caption{2d vortex-instanton solutions for various values of
$\kappa=1,16,1/16$. (a) shows the radial profile function $f$ and (b)
the Euclidean Lagrangian density $\mathcal{L}_E$.
The action of the solutions are indeed finite and are calculated
numerically to be 
$S_E(1)=\pi^3\mathsf{m}_3\mathsf{m}_4\times 4.658$,
$S_E(16)=\pi^3\mathsf{m}_3\mathsf{m}_4\times 5.241$ and
$S_E(1/16)=\pi^3\mathsf{m}_3\mathsf{m}_4\times 4.328$. } 
\label{fig:vortex}
\end{center}
\end{figure}

In Fig.~\ref{fig:vortex} are shown the profile function $f$ and the
Lagrangian density for various values of $\kappa=1/16,1,16$.
Vortex solutions in this model are somewhat similar to 
those in the Skyrme
model \cite{Gudnason:2014gla,Gudnason:2014hsa,Gudnason:2016yix} with a
similar potential term, albeit the kinetic term has a nontrivial field
dependence as in $K$-theories, see
e.g.~\cite{Babichev:2007tn,Adam:2008rf,Bazeia:2012uc}.

Notice that although there are two potentially logarithmically
divergent terms in the action, they come with coefficients
$\kappa^{-1}$ and $-\kappa^{-1}$ and thus cancel, leaving the action
integral convergent. 
Let us however examine the potential divergence in more detail.
The asymptotic behavior of the profile function is
\beq
f \sim \frac{\pi}{2} - A e^{-\sqrt{\kappa}r},
\eeq
where $A\in\mathbb{R}_{>0}$ is an undetermined constant.
Substituting into the Euclidean Lagrangian
density \eqref{eq:L2dvortex}, we get to leading order
(i.e.~approaching zero the slowest) at asymptotically large $r$:
\beq
\mathcal{L}_E \sim 2 A^2 e^{-2\sqrt{\kappa}r}
\left(1 + \frac{1}{\kappa r^2} + \frac{\kappa}{r^2}\right)
+ \mathcal{O}\left(e^{-4\sqrt{\kappa}r}\right),
\eeq
which leaves the action integral convergent, as promised.

\section{Discussion and conclusion}\label{sec:discussion}

In this paper we have constructed solitons in the supersymmetric
Skyrme model in three different codimensions from two through four.
The results are summarized in Table. \ref{tb:instantons}.
%%%%%%%%%%%%%%%%%%%%%%
\begin{table}[tb]
\begin{center}
  \begin{tabular}{|l||l|l|} 
\hline
Dim. & Instantons & Relevant terms \\ 
\hline \hline
4 & Pure Skyrme-instanton & Pure Skyrme term, no kinetic term  \\
3 & Skyrmion-instanton & Kinetic and Skyrme terms \\
2 & Vortex-instanton & Kinetic, potential and Skyrme terms \\ 
\hline
  \end{tabular}
\end{center}
\caption{Instantons in various dimensions.}
\label{tb:instantons}
\end{table}
%%%%%%%%%%%%%%%%%%%%%%
As mentioned, the supersymmetric Skyrme model lacks a kinetic term
because the auxiliary field equation eliminates it, leaving only a
potential term for the quasi-NG bosons behind.
In this paper, we concentrated on the theory restricted to the NG 
submanifold only, for which the supersymmetric Skyrme term is exactly
equal to the standard bosonic Skyrme term.
First we reviewed Speight's instanton in the 4-dimensional Euclidean
pure Skyrme model.
Then we performed a Scherk-Schwarz dimensional reduction to 3 dimensions
in which we constructed a 3-dimensional instanton that looks like a
squashed sphere.
Finally, we performed yet another anisotropic SS dimensional reduction
to 2 dimensions, in which we constructed a Euclidean vortex-instanton 
with finite action.
We would like to point out that this is -- to the best of our
knowledge -- the first global vortex with finite tension.

As mentioned above, the existence of a stable soliton necessitates the
existence of a pressure term. Two common terms are the kinetic term
and the potential term.
One possibility is to use the potential that is induced by the kinetic
term; we mentioned that it does not provide a potential for the NG
bosons, but only for the quasi-NG bosons: this is only if
$G$-invariance (here it is $\mathrm{SU}(2)\times\mathrm{SU}(2)$) is
kept. If we sacrifice the $G$-invariance, then we can make a
$G$-symmetry breaking potential in the theory. Another possibility,
yet to be explored, is to include a superpotential in the theory and
solve the auxiliary field equation again \cite{Sasaki:2012ka}. 

A comment on supersymmetry of the models is in order.
We have performed SS dimensional reduction on the pure Skyrme model
which is just a truncation of the supersymmetric Skyrme model to the
NG subspace. 
Although the solution of the auxiliary field has quite a non-linear
form on the non-canonical branch \cite{Gudnason:2015ryh}, it is
possible to write down the full untruncated Lagrangian and perform the
SS dimensional reductions to three and two dimensions. 
We pointed out that the 4d Skyrme-instantons by Speight are not BPS in 
the supersymmetric model. 
This is obvious because of the fact that the topological charge 
is ${\mathbb Z_2}$ and hence not ${\mathbb Z}$.
Consequently, two instantons annihilate each other and 
an anti-instanton is the same as the instanton itself. 
This is in contrast to BPS solitons for which two solitons have 
the mass exactly equal to twice the mass of a single soliton. 
Indeed, it is also possible to show that all the instantons discussed
in this paper are non-BPS solitons.
A brief discussion on the BPS properties of the instantons in three 
and two dimensions is found in appendix.

In this paper we have performed SS dimensional reductions,  
which can be obtained as 
the small compactification limit 
of the compactified circle with twisted boundary conditions (TBC).
This reduction can give a relation between 
solitons in different dimensions. 

For instance, 
this has been considered in Yang-Mills theory in
$\mathbb{R}^3\times S^1$
\cite{Lee:1997vp,Kraan:1998pm,Kraan:1998sn,Weinberg:2006rq},
by which an instanton (caloron) 
is decomposed into a set of BPS monopoles.  

Skyrme chains have been constructed in $\mathbb{R}^2\times S^1$ also
with TBC, which for chains of 1-Skyrmions are well-approximated by
the holonomy of Yang-Mills calorons \cite{Harland:2008eu}. 
Instantons in the principal chiral model were constructed on
$\mathbb{R}^2\times S^1$ with TBC
in \cite{Nitta:2014vpa,Nitta:2015tua}. 
A 3d instanton is first interpreted as 
a vortex ring with a U(1) modulus twisted once 
\cite{Gudnason:2014gla,Gudnason:2014hsa,Gudnason:2016yix},
and in the small compactification limit, 
it is decomposed into a vortex and an anti-vortex 
with the U(1) modulus twisted half (having half Skyrmion charges) 
\cite{Nitta:2014vpa,Nitta:2015tua}. 
The vortex-instanton in this paper is somewhat similar to 
these two cases.

Lumps (sigma model instantons) in the ${\mathbb C}P^n$ model in
$\mathbb{R}^1\times S^1$ with TBC were constructed in 
\cite{Eto:2004rz,Eto:2006mz,Eto:2006pg}.
A lump can be interpreted as a domain wall 
ring with the U(1) modulus twisted once,
and in small compactification limit 
it is decomposed into a kink and an anti-kink 
with U(1) modulus twisted half (having half lump charges) 
\cite{Eto:2004rz,Eto:2006mz,Eto:2006pg}.\footnote{
This was used to construct bions, 
a pair of a kink-instanton and an anti-kink-instanton 
with zero total instanton charge (exact bion solutions are available
for the $\mathbb{C}P^2$ model  
\cite{Dabrowski:2013kba,Misumi:2016fno}).
The application of bions to resurgence has been extensively studied 
\cite{Dunne:2012ae,Dunne:2012zk,Misumi:2014jua}.
}

These relations should hold for our 4d instanton and 3d instanton.
In four dimensions, a 3d Skyrmion-instanton is a string 
with SU(2) moduli.
When we make a ring the SU(2) moduli should be
 twisted somehow to induce a $\pi_4$ charge. 
 This was discussed in the context 
 in Helium-3; 
 a Shanker monopole string (characterized by 
 $\pi_3[\mathrm{SO}(3)]\simeq\mathbb{Z}$) is twisted to make a ring 
 producing an instanton 
 (characterized by $\pi_4[\mathrm{SO}(3)]\simeq\mathbb{Z}_2$)
 \cite{Nakahara:1986kn}.
Difficulties, however, are that SU(2) moduli cannot be uniquely
twisted along $S^1$ and that the $\pi_4$ charge is $\mathbb{Z}_2$ 
so twisting twice should be equivalent to untwisting.

A story, similar to the one in this paper, plays out in the
supersymmetric baby Skyrme model
\cite{Adam:2011hj,Adam:2013awa,Nitta:2014pwa,Bolognesi:2014ova,Queiruga:2016jqu},
where the kinetic term also vanishes when a nontrivial solution of the
auxiliary field equation is used \cite{Adam:2013awa}. Analogously  
to this paper, Scherk-Schwarz dimensional reduction can be carried out
also in that case, yielding a kinetic term and possibly 
a domain wall solution \cite{Eto:2004rz,Eto:2006mz,Misumi:2014jua} if
we perform it twice.

\subsection*{Acknowledgments}

S.~B.~G.~thanks the Recruitment Program of High-end Foreign
Experts for support.
The work of S.~B.~G.~was supported by the National Natural Science
Foundation of China (Grant No.~11675223).
The work of M.~N.~is supported in part by a Grant-in-Aid for
Scientific Research on Innovative Areas ``Topological Materials
Science'' (KAKENHI Grant No.~15H05855) and ``Nuclear Matter in Neutron
Stars Investigated by Experiments and Astronomical Observations''
(KAKENHI Grant No.~15H00841) from the the Ministry of Education,
Culture, Sports, Science (MEXT) of Japan. The work of M.~N.~is also
supported in part by the Japan Society for the Promotion of Science
(JSPS) Grant-in-Aid for Scientific Research (KAKENHI Grant
No.~25400268) and by the MEXT-Supported Program for the Strategic
Research Foundation at Private Universities ``Topological Science''
(Grant No.~S1511006).
The work of S. S. is supported in part by Kitasato University
Research Grant for Young Researchers.

\appendix

\section{BPS property of instanton solutions}

We remark that solitons in supersymmetric field theories are
potentially BPS states which preserve fractions of supersymmetry. 
BPS states in 4-dimensional supersymmetric higher-derivative theories  
have been classified in \cite{Nitta:2015uba}.
The BPS states satisfy the condition that the supersymmetry variation
of $\psi_M$, the fermionic partner of $M$, 
vanishes for some $\xi$ and $\bar{\xi}$:
\begin{align}
\delta (\psi_M)_{\alpha} = \sqrt{2} i (\sigma_{\text{E}}^{\mu})_{\alpha
 \dot{\alpha}} \bar{\xi}^{\dot{\alpha}} \partial_{\mu} M + \sqrt{2}
 \xi_{\alpha} F_M = 0,
\end{align}
where $\alpha = 1,2$ and $\sigma^{\mu}_{\text{E}} = (i \vec{\tau}, \mathbf{1}_2)$, 
$\bar{\sigma}^{\mu}_{\text{E}} = (-i \vec{\tau}, \mathbf{1}_2)$ which satisfy
$\{\sigma^{\mu}_{\text{E}}, \bar{\sigma}^{\nu}_{\text{E}} \} = 2
\delta^{\mu \nu} \mathbf{1}_2$ are the sigma matrices in Euclidean
space. 
More explicitly, the variation is found to be 
\begin{align}
\delta \psi_M =& \ \sqrt{2}  
\left(
\begin{array}{c}
(\partial_3 - i \partial_4) M \bar{\xi}^{\dot{1}} + (\partial_1 - i
  \partial_2) M \bar{\xi}^{\dot{2}} - i \xi_1 F_{M} \\
(\partial_3 + i \partial_4) M \bar{\xi}^{\dot{2}} +  (\partial_1 + i
 \partial_2) M \bar{\xi}^{\dot{1}} - i \xi_2 F_{M}
\end{array}
\right), 
\notag \\
\delta \bar{\psi}_M =& \ - \sqrt{2} i 
\left(
\begin{array}{c}
(\partial_3 + i \partial_4) \bar{M} \xi_1 + (\partial_1 - i
 \partial_2) \bar{M} \xi_{2} + i \bar{\xi}^{\dot{1}} \bar{F}_{M} \\
- (\partial_3 - i \partial_4) \bar{M} \xi_2 + (\partial_1 + i
  \partial_2) \bar{M} \xi_1 + i \bar{\xi}^{\dot{2}} \bar{F}_{M}
\end{array}
\right).
\label{eq:SUSY_variations}
\end{align}
We stress that the supersymmetry transformation parameters $\xi$ and
$\bar{\xi}$ are independent of each other in Euclidean spaces.

We are interested in codimension-four instanton-like configurations. 
For the non-canonical branch, we have $F_M, \bar{F}_M \neq 0$.
Let us consider the following 1/4-BPS condition
$\bar{\xi}^{\dot{1}} \neq 0$,
$\bar{\xi}^{\dot{2}} = \xi_1 = \xi_2 = 0$, then we have
\begin{align}
(\partial_3 - i \partial_4) M = (\partial_1 + i \partial_2) M = 0,
 \qquad \bar{F}_M = 0.
\end{align}
The last condition together with the equation of motion of the auxliary
field \eqref{eq:auxiliary_eom} with $f_{\pi} = 0$ implies the
following condition: 
\begin{align}
0 = \text{Tr} F_M F^{\dagger}_M = \text{Tr} \partial_{\mu}
M \partial_{\mu} M^\dag.
\end{align}
The only solution to these conditions is $M = \text{const}$.
The other choices of nonzero component of $\xi$, $\bar{\xi}$ result in
the same condition.
The other possible 1/4-BPS combinations of $\xi, \bar{\xi}$,
for example $\bar{\xi}^{\dot{2}} -i \xi_1 = \bar{\xi}^{\dot{1}}
= \xi_2 =0$, lead to the condition
\begin{align}
(\partial_1 - i \partial_2) M = F_M, \quad 
(\partial_3 + i \partial_4) M = 0.
\end{align}
This implies that $M$ is a holomorphic function of $z = x^3 + i x^4$
and its dependence in the $x^1,x^2$ plane is determined by the source
term $F_M$ on the right-hand side in the first equation. This is just
a vortex-lump type configuration discussed in \cite{Nitta:2015uba}.
The combinations like
$\bar{\xi}^{\dot{2}} - i \xi_1 = \bar{\xi}^{\dot{1}} - i \xi_2 = 0$
give lump-lump type configurations.
This is true even if we restrict the model to the NG subspace.
Therefore, there are no non-trivial 1/4 BPS instantons in
the pure Skyrme model in four dimensions.

For the 1/2-BPS condition, if we keep two of the four parameters,
it inevitably leads to the condition $F_M=0$ which results in a vacuum 
condition of $M$.
If we combine, for example, $\bar{\xi}^{\dot{2}} = i\xi_1$,
$\bar{\xi}^{\dot{1}} = i\xi_2$, this leads to the condition that $M$ is
independent of $x^3,x^4$.
Therefore codimension-four solitons are inconsistent with the 1/2 BPS
condition. 
We note that this is the story for codimension-four solitons.
We can, however, find 1/4 and 1/2 BPS configurations of codimension
two on the non-canonical branch \cite{Nitta:2015uba}.

The conclusion is that the pure Skyrme-instantons constructed by
Speight is not BPS in the supersymmetric Skyrme model.
In hindsight, it is obvious that they cannot be BPS since they carry
$\mathbb{Z}_2$ charge.
Although it is not a BPS solution in the sense that it does not
preserve fractions of supersymmetry, the instantons by Speight in four
dimensions do satisfy the equation of motion and saturate an energy
bound on the NG subspace. 

The same analysis can be applied even to the lower-dimensional models.
In lower dimensions, the supersymmetry variation of fermions is given 
by eq.~\eqref{eq:SUSY_variations} in which the derivatives in the
compactified directions $x^3,x^4$ are replaced by Killing vectors $G$ 
associated with isometries in the target space. 
Namely, the derivative with respect to the compactified direction induces
a motion along the Killing vectors, e.g.~$\partial_3 M = m G (M)$
where $m$ is a mass parameter.
These Killing vectors generate central
charges in the supersymmetry algebra.
Using this fact, it is obvious that the above discussion in four
dimensions holds true in 
lower dimensions. We find that only possible BPS instantons are
two-dimensional 1/2 BPS lumps in the massless theory, which are
characterized by a holomorphic dependence on $x^1,x^2$.
One finds that the Skyrmion-instanton and the vortex-instanton
discussed in the main body of this paper do not belong to this class
and they are thus non-BPS solitons.

\end{document}